# Why 6G?: Motivation and Expectations of Next-Generation Cellular Networks

Sudhir K. Routray, Addis Ababa Science and Technology University

Sasmita Mohanty, University of Aveiro

*Since the 1980s, the world has witnessed new mobile generations every decade. Each new generation is better than the previous in some ways. The recently emerging generation, 5G has several advanced features. However, it is doubted that there will be several short comings of this generation when compared with the other contemporary ICT alternatives. These short comings are going to be the main motivation for the next new mobile generation. According to the existing trends, this new version will be known as the Sixth Generation of Mobile Communication (6G). In this article, we show the main driving forces behind 6G, its expected features and key technologies are also been discussed.*

## Introduction

Mobile communication was started in the early 1980s. Since then every decade, we have observed new generations of mobile communication technologies with much improved features and performances. According to the observed trends, a new generation arrives in each ten years with their incremental versions found in between. For instance, second generation (2G) came in the 1990s, third generation (3G) in 2000s, fourth generation (4G) in 2010s and the fifth generation (5G) will hit the markets in 2020s. Intermediate versions were observed in the form of $A.bG$ which are normally better than $AG$ and inferior than $(A+1)G$. Normally, $A$ in a positive integer between 1 and 5, and b is an integer smaller than 10.

Currently, the Sixth Generation (6G) is a very new topic and its research is very much at its infancy. There are very few literatures available on this topic. In [1], beyond 5G networks are discussed. The main vision and requirements for 6G have been presented [1]. There are several possibilities beyond 5G in terms of new networks and services. In [2], several flexible radio access technologies beyond 5G have been analyzed. Specifically, the waveforms, frame design strategies and numerology of the beyond 5G radio technologies have been emphasized in this paper. High throughput satellites have generated a lot of interest in the recent years. In [3], a flexibility analysis model for high throughput satellites has been presented which would be required for the future generation flexibility related applications. In [4], consumer electronics driven global trends of communications have been presented. It shows the scale and pace of changes happening these days. In [5], the technical specifications and performance metrics of 4.5G (LTE Advanced Pro) have been presented and compared with the 5G specifications of ITU. It shows the intensity of current competition in the recent years. Optical communication is the trend setter in the high speed regime since the 1970s. The trends of highest available speeds through optical fiber are explained using Keck's law. In [6], the ways to maintain the high speed trends of Keck's law have been presented. This indicates that there is no real competitors of optical fibers as far as the data rates are concerned. 5G specifications were finalized by ITU in 2015. It has several improvements over the previous 4G mobile communication systems. In fact, 5G is much more advanced than the previous generations /versions of mobile communications. Several advanced networks are envisioned for the future Internet. The Bell Lab's perspectives on future networks have been shown in [8]. Light can be used for fast communications in the access area which is now known as Light Fidelity (LiFi). It has the potential to communicate at speeds of 100 Gb/s and beyond [9]. There are several access technologies available which are faster than 5G such as terabit passive optical network (PON), and wireless optical communication networks. Terabit PON is among the fastest access technologies currently known. Recently, terabit digital subscriber lines (DSLs) have been proposed for cheap and fast local area connections.

The reminder of this article is organized as three sections. In the next section, we present the main motivation for 6G. In the following section, we show the general trends of mobile generations which explain the genesis of new generations. In the last section we predict the nature, timeline and the social significance of 6G.

## Main Motivation for 6G

Motivation for every new mobile generation comes from the short comings of the previous generation. In case of 6G too, it is going to be the major reason. 5G has been specified by ITU with several advanced features. In Table 1, we show the main features of 5G. All these features are not available yet.

The current research achievements in the laboratory are also shown in Table 1. Those features will be improved over time in the coming years. Despite having all those advanced features, 5G will not be the perfect technology to meet all the user demands in terms of quality and quantity. Those short comings will be addressed in 6G.

*Table 1* ITU Technical Specification for 5G (Performance Indicators, their average values) their current laboratory achieved value and their special conditions for achieving the laboratory values

ICT restructuring is a continuous process. It happens at every company and country level. As a part of restructuring, technology and management related changes are added to the service provisioning every year. These changes provide strategic edges to the operators over their rivals. Therefore, irrespective of the state of the technologies, ICT restructuring will keep on adding advanced features to their services.

Once the 5G is rolled out in the major cities around the world, its performances can be measured directly. User experiences and 5G utilities in the real world can be assessed directly. This is expected to happen around 2020. It is true that all the requirements and expectations of 5G will not be fulfilled in 2020. Only the incremental advances will be observed over the years. The full potential of 5G can only be achieved after some years of its initial deployment. In Figure 1, we show the average value of the initial roll out specifications of 5G along with the average ITU specifications. For instance, the average ITU specification of data rate is 10 Gb/s (with peak data rate 20 Gb/s). However, it is expected that the best possible data rate in 2020 roll out version would be around 5 Gb/s. Similarly, the other performance parameters too remain much inferior to the ITU specifications.

*Figure 1* ITU Specifications of 5G vs. the best practical roll out specifications of 5G in 2020. The initial roll out specifications of 5G in 2020 is going to be much inferior to the ITU specifications of 5G.

Inherently, there are several drawbacks in the 5G development process in achieving the ITU specifications. For instance, software defined networking (SDN) is going to be an integral part of 5G. It would work through network function virtualization (NFV). SDN would slice or segregate different types of service allocations in to different categories known as slices. These slicing mechanisms would slowdown the speed or in other words the data rates will be affected by SDN. Similarly, clouds and fogs will have enhanced roles in 5G. Too many transactions or computing in clouds and fogs would increase the latency inherently. So, the latency reduction will be affected through the intense cloudification of 5G. All these drawbacks of 5G will take time to be solved. That is how a new generation would be brought in to advance the performances of mobile communication and computing. Security aspects in 5G are still not very strong. This is a big drawback for 5G.

From the performances of the previous generations, it has been observed that the mobile generations are not the best access technologies of their time. Some of their contemporary technologies outperform them. In case of 5G it is also true. By 2025, 5G is expected to be widely deployed. At that time there would be several better technologies in the access area. From the current technology trends, we can easily predict a few technologies in the access area which are better than the 5G. In the fourth section we show these technologies. These technologies would motivate for a better next generation cellular communication networks.

**The General Trends of Mobile Generations**
The general trends of development of mobile technologies follow a specific pattern. In order to understand any new generation we should go through the genesis of the past generations. It provides logical reasons behind the development of any new version or generation. Mobile communications started as a digital technology in the mid eighties and the commercial popularity around the world came in the early nineties. The 2G services were basically voice services. Gradually, new features were added to it such as the SMS service which is known as 2.5G. Basically it was value addition to the 2G services. Gradually, the data rate demands were on the rise. It was not possible in the typical GSM framework of that time. Therefore, for the faster communication of voice, data, and videos a new generation was required. Mobile companies (operators, vendors and associates) around the world joined together for this new generation. They formed a new group called Third Generation Partnership Project (3GPP) for the future long term evolution (LTE) of mobile technologies. These 3GPP initiatives streamlined the development of future mobile generations.

The 2G difficulties were resulted in the planning for 3G. However, the 3G specifications were not achieved directly. It was developed through several sub-3G versions such as 2.8G and 2.9G. Gradually, the 3G specifications were achieved as the technology for data and video communication he mobile platform got mature. In the 3G initiative, data transfer became very popular. Mainly the Internet access over the mobile phones was very attractive among users. Internet penetration after the arrival of 3G increased significantly. In a few years, the major demerits of 3G came to light. The users and the operators realized that the speed of communication over 3G are not realistic as the other contemporary technologies were much better than 3G. Both Wireless Fidelity (WiFi), and Worldwide Interoperability for Microwave Access (WiMAX) were much better than 3G in terms of data rates. WiFi was easily accessible in the static environment and was also cheaper in price when compared with the cellular services. WiMAX was much better than 3G in both the static and mobile conditions. This was main motivation for 3GPP to look beyond 3G.

Therefore, 4G was proposed whose performances were several times better than the 3G.

4G was developed gradually through the incremental versions of 3G. They were better than 3G and inferior to the 4G. In 2009, 4G roll out was started in a few cities. The widespread 4G deployment was achieved only in the second half of 2010s. The performances of 4G were way better than the 3G services but worse than WiFi and several other access technologies. Thus the next advanced version, 5G was planned by 3GPP. However, this time the plan was very much different. It was a giant leap over the previous versions. The 5G was proposed to be much superior to 4G in several aspects. The 5G specifications were mainly imitating the performances of optical access technologies. Passive optical networks (PONs) can provide hundreds of Gb/s data rates. Even as early as 2011, terabit PONs were tested in the laboratories. When compared with the PONs, wireless communications were far behind. One of the main aims of 5G was to provide fiber like experience in the wireless domain. If we follow the existing trends, 6G will be developed through the incremental versions of 5G.

## What 6G will be?

It is really very difficult to predict what exactly 6G will be. For instance the research on 5G started around 2009. However, the exact features of 5G were defined by the ITU in collaboration with other standardization bodies much later in 2015. Similar uncertainties were also observed for the predictions over 3G, and 4G at the time of their inception. Unless the complete standards are agreed by the expert participating groups around the world, it is not possible to predict the exact features. However, from the past experiences, we can predict the following features of 6G.

1. 6G will be better than 5G in all performance related aspects.
2. The peak data rates proposed for 5G is 20 Gb/s and the peak wireless data rates of 6G will be around 100 Gb/s. In fact, for 6G data rate will not be the main attraction. Rather, overall quality of performances would matter.
3. The 5G latency of 1 ms would be further reduced to µs order.
4. Device densities and IoT connectivity will be much denser than the 5G. The real digital ecosystem would be realized in 6G.
5. Energy efficiency will be certainly better than the 5G.
6. Spectral efficiency will be better than that of 5G.
7. More artificial intelligence (AI) and machine learning applications will be found in 6G. In fact, mobile handsets would not be used just for communication; rather they would become personal digital assistants of individuals.
8. 6G will be a hybrid ubiquitous network which would incorporate all advances contemporary technologies in its ambit. In 6G, optical networking would play bigger roles than any previous generations.
9. Security is expected to be much better in 6G when compared with 5G. Quantum communication and cryptography are expected to be deployed in 6G.
10. More complex device to device communications are expected in 6G.

6G would be a very complex hybrid system where optical and wireless communications will have equally important roles. 6G would incorporate all the advanced ideas and techniques of the contemporary technologies. Currently, the following communication technologies are better than the proposed 5G specifications.

1. PONs provide higher data rates, lower latency and better reliability
2. Even terabit PONs are available which can carry gigantic amounts of data
3. Terabit Ethernet will be available shortly. 400 Gb/s Ethernet is already available.
4. High throughput satellites (HTSs) are capable of dealing with high data rates. For high data rate communication over the oceans, deserts and unconnected territories it is essential.
5. LiFi communication in the access area can provide data rates of 100 Gb/s and beyond.

It is expected that 6G would incorporate all these technologies when deployed.

### Who are Interested for 6G?

5G specifications are good enough for individual needs. Thus very few individuals would be interested for 6G. There are special interest groups who want the next mobile generation. First of all, the telecommunication vendors are the ones who benefit the most from the new mobile generations. They are the experts who design and develop the networks for new generations. Without new mobile generations, their businesses do not grow fast. So, it is always a win-win condition for the telecom vendors to roll out a new technology. The new technologies come with new techniques and equipments which are the direct source of revenue for them. The next beneficiaries are the telecom operators. Operators make money from their services. New generations come with new services and thus new income sources for them. Of course, it true that some of the old services also become obsolete or do not produce revenue. However, in the competitive business arena, the operators want to defeat their rivals through advanced technologies. Governments also want new technologies to earn new taxes and revenues.

### Timeline for 6G Development

Timeline for 6G is expected to follow the previous timeline patterns of previous generations. We observed a time gap of almost 10 years between two mobile generations starting from 2G to 5G. A similar time gap is expected to reach a preliminary version of 6G. As expected, 5G will not be rolled on all over the world in 2020. Rather only some cities across the world will have it. Around 2025, 5G will be deployed widely across the world. Rural deployment in the developing countries may take even longer. Thus, the incremental versions of 5G will be developed after 2025 which are expected to be better than 5G and inferior to 6G. This gradual process of innovation would make 6G ready for deployment around 2030. As it happened with the other generations, large scale deployment of 6G will not happen immediately. Rather, it will be adopted slowly. It may not be an attraction for many developing countries in 2030s as 5G itself would be too advanced for them. The individual communication demands can be easily met by the 5G specifications. Thus 6G and its subsequent versions may remain limited only to the business and high performance applications.

*Impact of 6G on Society and Business*

It is very difficult to predict how an advanced technology would affect the society. In the last two decades, we have witnessed several disruptions through the mobile technologies. Similar disruptions are equally possible through 6G. However, 5G is likely to do a lot of changes in the society. Therefore the arrival of 6G may not be a surprise for several social activities. AI and machine learning are going to play some roles in 5G. However, that will not be very mature in just one decade. Therefore, 6G is envisioned as the true AI powered mobile generation.

Large scale satellite and mobile conjugation is an expectation in 6G. It can enhance the communication performances over the oceans and on board the flights. Satellite conjugation with mobile infrastructure is not very feasible at the moment. It may get started in the 5G regime. However, it is expected that only in 6G, satellite and mobile infrastructure will be very much compatible. Through satellite conjugation the true ubiquitous communication will be possible.  IoT networks too will get a boost through the satellite connectivity. 6G is perhaps the right mobile technology to bring all these new applications and possibilities.

*The G-Race*

Since the 1980s, a G-race is observed in ICT innovation to bring forward a new mobile generation every decade. It is now clear that 5G will be rolled out in several cities around world in 2020. By 2025, this technology would be mature and almost every country in the world would be using it. Based on these observed trends, we expect the arrival of 6G in early 2030s or a little before 2030 as shown in the deployment timeline. However, the G-race may not be as intense as we found in the previous cases of 3G, 4G and 5G. That is mainly due to the advances achieved in 5G. We can compare this with the different generations of computer processors.

Initially, when the microprocessors came in the early 1970s, it was a revolution altogether. The second generation came in the late 1970s, followed by third generation in the early 1980s. The fourth generation of 32-bit processors in the late 1980s brought a lot of aura in the computing world. The fifth generation Pentium processors arrived in the 1990s which changed the computing paradigm to a large extent. Since then, processors are being updated to become faster and better. Even the generations have evolved as sixth, seventh, eighth, and beyond. However, the craze, aura, and publicity are no more observed since the fifth generation. No more the generation races are relevant in the microprocessor innovation. That is mainly due to the speed of the processors achieved in the fifth generation. The speed of the processors since the fifth generation is good enough to carry on most of the common personal and office works. The faster processors are required only for the special projects or advanced computing applications. Therefore, the craze for the faster new processors is no more like the ones seen in the 1990s. A similar saturation like effect is expected to be observed in the mobile generations. After 5G, the common users will not look for faster communication systems desperately. However, just like the microprocessors, some special purpose communication tasks may need the speeds and performances higher than that of 5G.

Overall, the G-race in mobile communication would be irrelevant after 5G. Though the new mobile technologies get developed under the G-hierarchy, they would not get the big attention from the public. However, it is possible that the ICT may develop in a whole new direction empowered by AI and machine learning.

## Conclusion

In this article, we presented the potential driving forces behind the development of 6G. The major short comings of 5G will lead to the next generation of mobile ICT. Computing capabilities of the 6G devices would be better than those of 5G. AI and machine learning applications would find prolific presence in 6G. The deployment of 6G would start around 2030. It would be very much energy and bandwidth efficient. The new dimensions such as quantum communication and satellite integration are expected to find places in the 6G ambit. The G-race in mobile communications would be over with 5G. So, 6G and other future generations will not create the craze and publicity like the previous versions.

## References

[1] K. David, and H. Berndt, "6G Vision and Requirements: Is There Any Need for Beyond 5G," IEEE *Vehicular Technology Magazine,* vol.13, no. 3, pp. 72-80, Sep. 2018.
[2] Z. E. Ankarali, B. Peköz, and H. Arslan, "Flexible Radio Access Beyond 5G: A Future Projection on Waveform, Numerology, and Frame Design Principles," *IEEE Access,* vol. 5, pp.18295-18309, 2017.
[3] K. Kaneko, H. Nishiyama, N. Kato, A. Miura, and M. Toyoshima, "Construction of a flexibility analysis model for flexible high-throughput satellite communication systems with a digital channelizer," *IEEE Transactions on Vehicular Technology,* vol. 67, no. 3, pp. 2097-2107, Mar 2018.

[4] S. Mohanty, and S. K. Routray, "CE-Driven Trends in Global Communications: Strategic sectors for economic growth and development," *IEEE Consumer Electronics Magazine*, vol. 6, no. 1, pp. 61-65, Jan. 2017.

[5] S. K. Routray, and K. P. Sharmila, "4.5G: A Milestone Along the Road to 5G," in Proc. of 5th IEEE International Conference on Information, Communication and Embedded Systems (ICICES), Chennai, Feb 25–26, 2016.

[6] S. K. Routray, A. Javali, R. Nymangoudar, and L. Sharma, "Latching on to Keck's Law: Maintaining High Speed Trends in Optical Communication," in Proc. of 4th IEEE International Conference on Advanced Computing and Communication Systems (ICACCS), Coimbatore, Jan. 2017.

[7] S. K. Routray, and K. P. Sharmila, "Green Initiatives in 5G," in Proc. of 2nd IEEE International Conference on Advances in Electrical, Electronics, Information, Communication and Bioinformatics (AEEICB), Chennai, Feb. 2016.

[8] M. K. Weldon, *The Future X Network: A Bell Labs Perspective*, CRC Press, 2016.

[9] H. Haas, L. Yin, Y. Wang, and C. Chen, "What is lifi?," IEEE/OSA *Journal of Lightwave Technology,* vol. 34, no. 6, pp. 1533-1544, Mar. 2016.

[10] N. Cvijetic, M. Cvijetic, M.-F. Huang, E. Ip, Y.-K. Huang, and T. Wang, "Terabit optical access networks based on WDM-OFDMA-PON," IEEE/OSA *Journal of Lightwave technology,* vol. 30, no. 4, pp. 493-503, Feb. 2012.

[11] J. M. Cioffi, K. J. Kerpez, C. S. Hwang, and I. Kanellakopoulos, "Terabit DSLs," *IEEE Communications Magazine,* vol. 56, no. 11, pp. 152-159, Nov. 2018.

[12] S. K. Routray, "The Changing Trends of Optical Communication," *IEEE Potentials Magazine*, vol. 33, no. 1, pp. 28-33, Jan 2014.

*Sudhir K. Routray* (sudhir.routray@aastu.edu.et) works as an associate professor in the Department of Electrical and Computer Engineering at Addis Ababa Science and Technology University, Addis Ababa, Ethiopia. He received his BE in Electrical Engineering from Utkal University, India; MSc in Data Communications from The University of Sheffield, UK; and PhD in Telecommunications from University of Aveiro, Portugal. He has more than 60 publications in journals, conferences and books. His areas of interest are: 5G, IoT, network science, and optical networking. He is a senior member of IEEE.

*Sasmita Mohanty* (sasmita@ua.pt) is a PhD scholar in the Department of Economics, Management, Industrial Engineering, and Tourism at the University of Aveiro, Portugal. She has an MSc degree in Management from University of Aveiro, Portugal. She received her Master's and Bachelor's degrees in Economics from Ravenshaw University, India in 2008 and 2006 respectively. She is a member of Internet Society and EAI (European Alliance for Innovation). She is a columnist and reviewer of some reputed magazines. Her areas of interests are: telecommunication economics, telecommunication management, strategic management, and managed services in telecommunications.